# Influence by proximity effect on ultrasound attenuation in Cu-Nb composite system at low temperatures.

V.O. Ledenyov, D.O. Ledenyov, O.P. Ledenyov and M.A. Tikhonovsky

*National Scientific Centre Kharkov Institute of Physics and Technology, Academicheskaya 1, Kharkov 61108, Ukraine.*

The attenuation of longitudinal ultrasonic wave with the frequency of *30 MHz* in *Cu-Nb* copper-niobium (*40* vol%) composite system at low temperatures from *0.35 K* up to *2 K* is researched. It was found that the ultrasonic attenuation decreases in a *Cu-Nb* multi-filamentary composite sample at low temperatures in distinction to the pure *Cu* copper or *Nb* niobium homogeneous bulk samples. It is well known that the contact between the normal metal *N* and the superconductor *S* is characterized by an appearance of superconducting properties in the thin surface layer of normal metal *N*, because of the presence of proximity effect at low temperatures *T*. This phenomenon is observed on the temperature dependent distance $\xi_N(T)$ in the normal metal *N* at the normal metal – superconductor *NS* boundary. It is assumed that the transition from the normal state *N* to the superconducting state *S* must be accompanied by the decrease of magnitude of ultrasonic wave's electronic energy absorption in normal metal *N*. The experimental results show that the superconducting electron coherence length in the normal thin layer is temperature dependent $\xi_N(T)$. The electron mean free path *l*, which is dependent on the impurity scattering in volume and near *Cu-Nb* interfaces, was measured by the three independent methods at low temperatures. It is understood that the Andreev reflections are present in *S-N-S* structure, making some influence on the measured ultrasonic longitudinal wave attenuation in *Cu-Nb* composite samples at ultra low temperatures. The measured experimental results of ultrasonic longitudinal wave attenuation in *Cu-Nb* composite samples at low temperatures are in good agreement with the theoretical modeling data, obtained in case of dirty local limit.

## INTRODUCTION.

The composite samples of copper with layered and multi-filamentary superconducting materials represent a typical superconductor-normal metal-superconductor (*S-N-S*) system, in which the *N-S* layered structure has a strong influence on the electronic and superconducting physical properties of a sample. The proximity effect [1] was observed in different *S-N-S* systems, and one of the manifestations of this effect is an appearance of superconducting energy gap in normal metal layer close to the *N-S* boundary. In the multilayered *S-N-S* (*Nb/Cu/Nb*) systems, the *Josephson junction* properties [2] were found to exist, and the *3D-2D* dimensional crossover [3,4] was observed experimentally, and their influences on the magnitude of superconducting critical current [5] and some other physical properties [6] of samples were researched comprehensively. In our research work, the *Cu-Nb* thin layer composite samples have been synthesized with the aim to research the physical properties of low temperature superconductors. The proximity effect was observed in the thick one layer *Cu-Nb* system at temperatures down to *21mK* in presence of diamagnetic properties in [7]. Until this moment, the ultrasonic attenuation, which is sensitive to the *N-S* transition in metals as it is shown in Bardeen, Cooper and Schrieffer (*BCS*) theory [8], was not measured in this type of layered composite samples, which are not the bulk materials. In the layered *S-N-S* systems, the influence by electron- phonon interaction on the Josephson electronic properties was first observed in the thermoelectric effect [9], and then in the acoustics-electrical effect [10]. The *S-N-S in situ* formed multi filamentary composite samples are more suitable for the investigation of ultrasonic attenuation [11, 12]. The physical properties of these materials are similar to those in thin multilayered films, hence, in this paper, we are going to present the results of low temperature ultrasonic attenuation measurements in *Cu-Nb* copper - niobium (40 vol%) *in situ* formed multi-filamentary composite samples. We would like to note that the mechanical behavior of similar materials was researched in a number of other research papers [11-13], and also their normal state electrical resistivity was measured in [14], and the superconducting properties were studied in [12]. The mechanical and superconducting properties of the composites with micro-filamentary *Nb* fibers in *Cu* matrices, prepared by the stack and draw method, were investigated in [15-17]. The critical magnetic fields $H_c$ in ferromagnetic–superconductor (*Fe/Pb*) thin films were measured by the acoustic surface wave attenuation in [18].

# EXPERIMENTAL MEASUREMENTS.

## A. *Cu-Nb* composite sample synthesis.

The *Cu-Nb* two-phase alloys, containing *Nb* (*40 vol%*), were prepared with the use of high-purity metals by their melting in high-purity graphite cast in pure helium atmosphere as in [13]. The ingots, which contained the *Nb* dendrites with the dimension close to *10* micrometers, were pressed to the diameter *20 mm*, and then drawn in several stages to the diameter *4.95 mm* with the intermediate annealing at temperature of *1020 K*. The samples with the *10 mm* and *30 mm* length were cut-off from this cylindrical rod by the electro-erosion method for the ultrasonic and normal-state resistivity measurements. The etched surface shows an uniform dense distribution of *Nb* filaments with the average dimension close to *3 μm*, and the average distance between the filaments in copper matrix of *~(3–4) μm*. The conditional pattern of a sort of crosscut of a cylindrical composite *Cu-Nb* sample is shown in Fig. *1*.

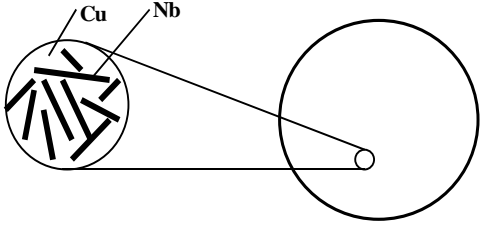

*Fig. 1. The conditional image of structure of filaments of niobium Nb on the cross section of a cylindrical composite Cu-Nb sample.*

The *Nb* filaments have the limiting length of a few hundreds *μm*, and randomly intersect each other. Their longitudinal orientation is in strong coincidence with the cylindrical rod axis.

## B. *Cu-Nb* composite sample properties.

The electrical conductivity of a composite sample with the *Nb* volume fraction $V_{Nb}$ is a linear combination of the conductivities of its components [16] in the case of negligible surface scattering effects. The average conductivity is given by the expression

$$\sigma = (1 - V_{Nb})\sigma_{Cu} + V_{Nb}\sigma_{Nb},$$

where $\sigma_{Cu}$ and $\sigma_{Nb}$ are the copper and niobium conductivities respectively. In our case, the volume fraction of the *Nb* niobium is equal to $V_{Nb}=0.4$, and the ratio $\sigma_{Cu}/\sigma_{Nb} \geq 10$ is true at room temperature, and the sample's conductivity is almost fully determined by the *Cu* copper conductivity. The electrical resistivity values $\rho_{293K}=1/\sigma_{293K}=2.13\cdot10^{-6}$ $\Omega cm$ and $\rho_{10K}=7.22\cdot10^{-8}$ $\Omega cm$ were measured by the four probes method, and the resistance ratio was approximately equal to *29.5*. The electrical conductivity in the simple electron theory [19] is defined as

$$\sigma = ne^2\tau/m^*,$$

where *n* is the electron density, *e* is the electron charge, *τ* is the relaxation time, *m\** is the effective electron mass. The slight modification of this expression may be presented as the equation

$$\rho l = p_F/ne^2,$$

where *l* is the electron mean free path in normal metal *N*, and it can be used for determination of the electron mean free path in normal metal *l* [20,21]. In the case of the *Cu* copper [19], $p_F = m^*v_F$, $m^*=1.38$ $m_e$, $n=8.45\cdot10^{22}$ $cm^{-3}$, $v_F=1.57\cdot10^8$ $cm/s$ for *T=10 K*, we found that the electron mean free path $l \approx 1.4\cdot10^{-4}$ $cm$. It was measured that the electrical resistivity $\rho=0$ below the *Nb* niobium critical temperature of *9.2 K,* and the sample is in a superconducting state, because in our case the niobium concentration is higher than the percolation limit for the disordered mixed system, therefore the superconducting current can flow through the *Nb-Nb* filaments contacts in a composite sample even when the *Cu* copper is in a normal metal state.

## C. Ultrasonic longitudinal wave attenuation measurements in *Cu-Nb* composite sample.

The ultrasonic attenuation was researched in the *Cu-Nb* composite samples with the length of *10 mm* and the diameter of *4.95 mm,* which have been prepared with high accuracy for precise ultrasonic measurements. The ultrasound propagated along the cylindrical sample axes in the direction of drawing process. The low temperature liquid $^3He$ cryostat with the adsorption cryopumping of $^3He$ vapor was placed in $^4He$ cryostat (see Fig. *2*).

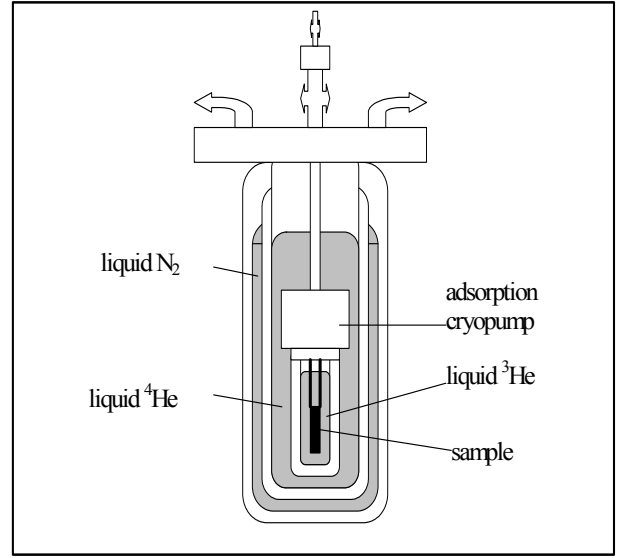

*Fig. 2. Scheme of a cryogenic part of measurement setup used for measurements of ultrasonic signal attenuation in composite multi-fibers Cu-Nb samples.*

The temperature was measured by the germanium thermometer and by the $^3He$ vapor registration with the precision $\pm 0.001\ K$. The sample was immersed in liquid $^3He$. The experimental ultrasonic attenuation measurement equipment was similar to the measurement setup, which was early used for the investigation of ultrasound attenuation in the intermediate state in superconductors [22] and the attenuation oscillations at low magnetic field in pure metal [23] (see Fig. 3).

The *30 MHz* signal generator generated the electromagnetic wave pulses of *1 μs* duration with the repetition frequency of *(50–200) c/s,* which have been transmitted by a coaxial cable to the *X*-cut quartz transducer with the *10 MHz* main tone mode frequency, which was mounted on one base of a composite sample. The signal from the second quartz transducer, attached to the other sample's end, has been transmitted to the receiver, and it was possible to compare the magnitude of experimental signal with the magnitude of pulses from the other high stable signal generator, which transmitted its signal to the precise attenuator, and then to the same receiver.

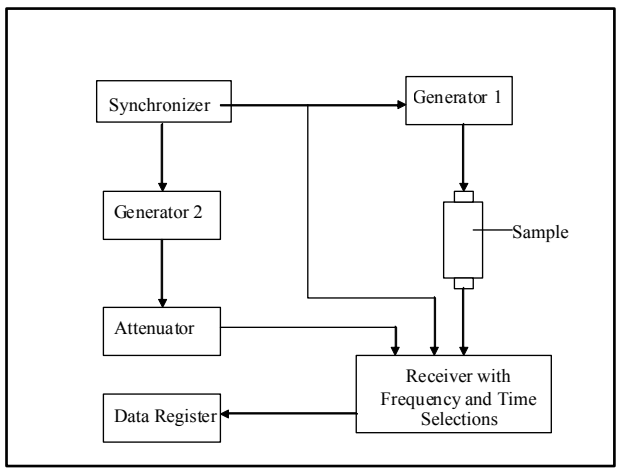

*Fig. 3. Block scheme of measurement setup for research of ultrasound absorption in a composite multi-fibers Cu-Nb sample at 30 MHz frequency.*

It was possible to measure the amplitudes of echo-pulse ultrasonic signals, which propagated in the *Cu-Nb* composite samples, using the time selection technique. The accuracy of ultrasound attenuation measurements in the *Cu-Nb* composite samples is $\pm 0.05\ dB$. It was found that the *Cu-Nb* composite samples have very advanced ultrasonic transmission properties with the near *20* echo-signals propagation without visible dispersion and with good exponential echo-signal attenuation dependence.

## RESEARCH RESULTS.

We measured the temperature change of attenuation difference between the first and fifth echo-signals with the acoustic signal path equal to *8cm* in the *Cu-Nb* composite samples. The longitudinal ultrasound velocity along the rod axis was $(5.11 \pm 0.02) \cdot 10^5\ cm/s$ as obtained from the pulse propagation time measurements. This velocity values are close to $5.09 \cdot 10^5\ cm/s$ for the ultrasound propagation direction *<110>* in the *Nb* niobium [24], and to $5.12 \cdot 10^5\ cm/s$ for the *Cu* polycrystal. The ultrasound propagation velocities for the different directions of *Cu* single crystal are discussed in details in [25]. The velocity in polycrystalline *Cu* copper for the ultrasonic longitudinal waves is $5.01 \cdot 10^5\ cm/s$ [26] at room temperature, and the *Cu* elasticity properties depend on its texture [26, 27]. In [11], it was determined that, after the deformation in *Cu-Nb in situ* formed specimens, the *Nb* bcc crystals develop the *<110>* fiber texture. In our composite samples, the *X*-ray analysis shows that the *Nb* filaments have this texture too. The low texture for the *Cu* matrix toward the *<111>* direction was observed. We propose, going from the ultrasound velocity measurements, that the *Cu fcc* matrix, after the deformation and annealing, has a polycrystalline structure with the texture slightly close to the <111> direction along the deformation axis. In our case, the ultrasonic longitudinal wave length was equal to $\lambda \approx 1.70 \cdot 10^{-2}\ cm$, and it was more longer than the intrinsic dimensions of *Cu-Nb* multi-filamentary composite sample structure $d \sim (3-7) \cdot 10^{-4}\ cm$, thus the sample can be considered as a homogeneous body without any dependence of the ultrasound velocity on the multi-filamentary structure periodicity, which may occur in sample with $\lambda / d \sim 1$ [28].

The temperature dependence of the *30 MHz* ultrasonic longitudinal wave attenuation in the *Cu-Nb* composite samples is presented in Fig. 4.

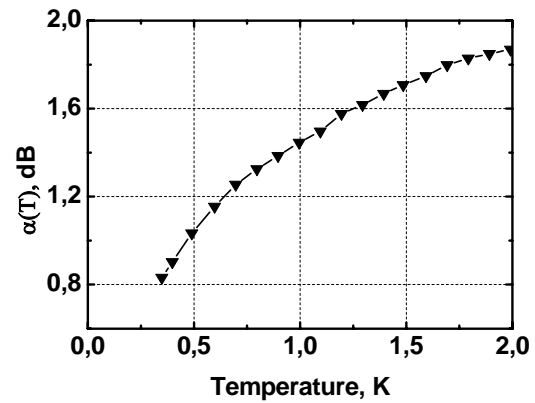

*Fig. 4. Temperature dependence of ultrasound absorption α(T) at 30 MHz frequency in Cu-Nb sample between 1 and 5 echo impulses.*

It demonstrates that the ultrasound attenuation nonlinearly decreases with the temperature decrease. The full ultrasonic longitudinal wave attenuation change in temperature range of *0.35-2.0 K* is equal to ~ *1.1 dB* or *0.138 dB/cm*. There are two possible phases in the researched *Cu-Nb* composite samples at ultra low temperature: the *Nb* superconducting filaments and the *Cu* normal metal matrix. According to the theory [8, 29], the ratio of ultrasonic longitudinal wave attenuation in superconducting state and in normal state is

$$\alpha_s / \alpha_n = 2/[1 + exp\{\Delta(T)/k_B T\}], \quad (1)$$

where $\Delta(T)$ is the temperature dependent superconducting energy gap, $k_B$ is the Boltzmann constant, and $T$ is the temperature. The ultrasonic longitudinal wave attenuation due to the interaction with the electrons in normal metals $\alpha_n$ as it was calculated in [30], and in other works (see, for example, the review [31]), and in the case of the *Cu-Nb* sample for $ql<<1$, it is equal to

$$\alpha_n = (\hbar q^2 / 4\pi^3 M v_s) \int D^2 l ds = (4/15) nmv(q^2 l)/Mv_s, \quad (2)$$

where $q = 2\pi/\lambda$ is the wave number, $M$ is the metal density, $v_s$ is the ultrasound velocity, $D$ is the deformation parameter, $\hbar = h/2\pi$ is the reduced Plank constant.

The ultrasonic longitudinal wave attenuation in the pure superconducting *Nb* niobium was researched in [24], and at the temperature below of *3 K*, its decrease should be negligible, because $\alpha_s/\alpha_n \to 0$. The ultrasound attenuation in the normal metal such as the *Cu* copper was measured in [25, 32], and it was found that, at the temperature $T<10\ K$, the magnitude of ultrasound attenuation is constant, because the electron mean free path $l$ doesn't depend on the temperature $T$. As we have shown above, the electron mean free path is equal to $l \approx 1.3 \cdot 10^{-4}$ *cm*, going from the electrical resistivity measurements in the *Cu-Nb* composite samples at the temperature of *10 K*. This value of the electron mean free path $l$ is less than the value reported in [32], where $l \approx 6.1 \cdot 10^{-4}$ *cm* at $T=30\ K$. In our researched *Cu-Nb* composite samples, the electron mean free path $l$ has to be independent on the temperature $T$ at the temperatures below $T<10\ K$. Note, that according to the measurements [11, 14], the *Cu* copper matrices may be more pure with $\rho \approx 2 \cdot 10^{-8}\ \Omega cm$. The calculations for the *Cu* copper give $\alpha_n \approx 7.16 \cdot 10^{-2}\ dB/cm$ at the frequency of *30 MHz* with the electron mean free path $l = 1.26 \cdot 10^{-4}$ *cm*. In our *Cu-Nb* composite samples with 60vol% *Cu*, the ultrasonic longitudinal wave attenuation should be equal to $\alpha_n \approx 3.44 \cdot 10^{-1}\ dB$ between the first and fifth echo signals. This ultrasonic longitudinal wave attenuation magnitude must be observed, when the thin layers of the *Cu* copper contact with the *Nb* niobium, and become superconducting, because of the proximity effect. The divergence in ultrasonic longitudinal wave attenuation levels, which was registered in our experiments, and calculated from the expression (2), was present due to:

i) the electron mean free path dependence on the temperature $T$, going from the Matthiessen's rule $l^{-1} = l_0^{-1} + l^{-1}(T)$, where $l_0$ is the mean free path attributed to the impurity scattering, $l(T)$ is the electron-phonon scattering mean free path with the $T^{-5}$ dependence;

ii) it is well known that there is a discrepancy between the $l$ obtained from both the electrical data and the ultrasound attenuation measurements data [31]. The electrical conductivity depends on an integer over the *Fermi surface* of mean free path $l$; and the ultrasound attenuation depends on an integral of the product $D^2 \cdot l$. In our case, the *Cu-Nb* composite samples are not very pure $ql<<1$, and the anisotropy of deformation parameter $D$ doesn't play a main role in electrical and ultrasonic results divergence. The total mean free path $l$, which is determined by the electrons scattering on the impurities located in volume and near the *Cu-Nb* boundaries is $l^{-1} = l_{0V}^{-1} + l_{0B}^{-1} + l^{-1}(0)$, where $l_{0V}$ is the mean free path in the *Cu* copper volume, and $l_{0B}$ is the mean free path associated with the boundary impurity. At the low temperature, the impurity located at *N-S* boundary doesn't take part in the electron excitations-impurity scattering process, when the superconducting energy gap appears near the *N-S* boundaries under the proximity effect action, and the *Andreev reflections* [33] can be observed. These electron excitations, which are located in normal metal region with the low value of induced superconducting energy gap, have more longer mean free path, interacting with the ultrasonic wave, and give their contribution to the ultrasound attenuation [34]. The ultrasound frequency $\omega$ is less than the frequency of electron-impurity collisions, and it is possible to consider the magnitude of electrical fields, created by the ultrasound wave as a constant along the mean free path $l$. The *Andreev electron-hole transformations* on the *N-S* boundaries don't have destructive influence on the ultrasound energy dissipation process as a result of interactions with the electron excitations [34]. Using the eq. (2) and the obtained experimental value of ultrasonic longitudinal wave attenuation $\alpha_n \approx 1.1\ dB$ for the *Cu-Nb* composite samples, we find that the mean free path $l$ in the *Cu* copper volume is $l_{0V} \approx 4 \cdot 10^{-4}$ *cm*. The results of the electrical measurements [11, 14], also, lead to the conclusion that the volume of the *Cu* copper matrix is more pure than the regions near the *Cu-Nb* boundaries, where the impurities concentration is big enough.

## DISCUSSION.

Due to the use of simple model of the *S-N-S* structure for the characterization of the *Cu-Nb* multi-filamentary composite samples [14] with the *Nb* niobium filaments thickness $d_S$, and with the distance between the filaments in *Cu* copper matrix $d_N$, it is possible to calculate the temperature dependence of ultrasonic longitudinal wave attenuation magnitude ratio $\alpha_s/\alpha_n$ for the *Cu* copper matrix, and to compare it with the experimental dependence in Fig. 5.

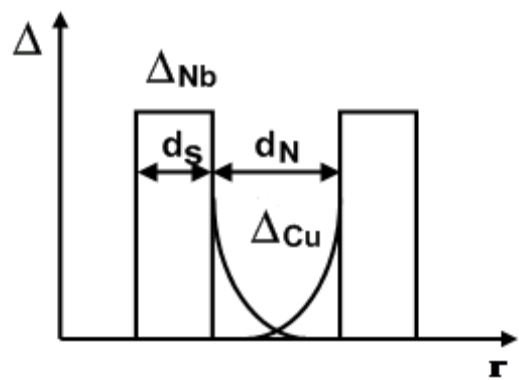

*Fig. 5. Conditional chart of magnitude of $\Delta$ superconducting energy gap in Nb niobium and Cu copper layers.*

According to the proximity effect theory [1] the superconducting order parameter can be expressed as $\Psi(r,t) \sim \Psi_0 \exp[i(pr - \varepsilon t)/\hbar]$, where $p$ is the electron pair momentum, $\varepsilon$ is the energy. It has an exponential decay in the normal metal as $\Psi(x) \sim \exp[-kr_N]$, and $r_N$ is the distance from the N-S boundary, $k = 2\pi/\xi_N$, $\xi_N$ is the superconducting coherent length in the normal metal. Theory [1] gives the temperature dependence of the $\xi_N$ for the two cases, when the normal metal is pure and electron mean free path $l > \xi_N(T)$:

$$\xi_N(T) = hv/2\pi k_B T,$$

and when the normal metal is dirty and $l < \xi_N(T)$:

$$\xi_N(T) = (hvl/6\pi k_B T)^{1/2}.$$

In this approach, the superconducting order parameter, which is induced into the normal metal, because of close contact of the normal metal with the superconductor, has both the distance and temperature dependences

$$\Psi_N(r_N, T) \sim \Psi_{N0} \exp[-r_N/\xi_N(T)].$$

In the case of local limit (dirty metal), it is possible to propose that the ultrasound attenuation is determined in each point of normal metal by the value of energy gap $\Delta(x,T)$, which is proportional to the local value of $\Psi_N(x,T)$ [35], and the full ultrasound attenuation is an integer over the distance $r_N$ from the N-S boundary with the limits from $0$ up to $d_N/2$. Using the dimensionless distance $x = r_N/(d_N/2)$, we may write the expression for the symmetrical units of the structure

$$\alpha_S(T)/\alpha_N = \int_0^{d_N/2} \frac{2}{1 + \exp[\frac{A\exp(-r_N/\xi_N(T))}{k_B T}]} dr_N =$$

$$= \int_0^{0.5} \frac{2}{1 + \exp[\frac{A/k_B}{T \exp(BT^{1/2}x)}]} dx,$$

where $A$ is a constant, which is proportional to the energy gap value on the N-S boundary $\Delta_{N0}$, $B$ is a ratio $d_N/2\xi_N (T=1 K)$, which is approximately equal to $\approx 0.75$. This dependence is shown in Fig. 6.

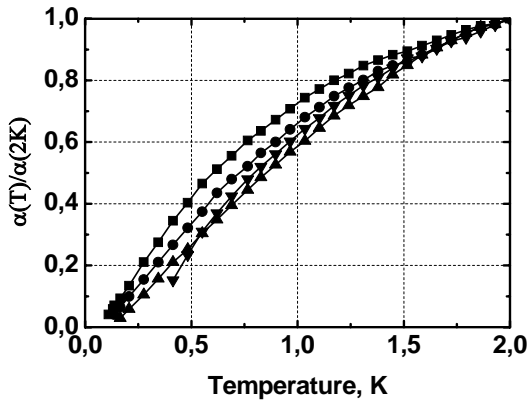

*Fig. 6. Calculated curves of the ratio of $\alpha(T)$ ultrasound attenuation at temperature T to the ultrasound attenuation at T=2K for the energy gap magnitudes: $\Delta_{N0}=36mK$ (square), $\Delta_{N0}=57mK$ (circle), $\Delta_{N0}=95mK$ (up triangle) and experimental curve (down triangle)*

The comparison of experimental results with the calculated theoretical modeling predictions indicates that this expression makes a good approximation of the ultrasound attenuation in Cu-Nb multi-filamentary composite samples at presence of the proximity effect. Using the expression for the constant value $A$, which is close to the *Bardeen Cooper Schrieffer* (BCS) theory expression for the energy gap $A = 3.5 k_B T^*$, where $T^*$ is the characteristic temperature, we found the $T^* \approx 70$ mK in temperature range of (2-0.6) K, and for the temperatures below 0.5 K, for example $T^* \approx 100$ mK and below. In [7], it was found that there is the characteristic temperature $T^* \approx 40$ mK for the Cu layer, which covers the Nb, and this temperature plays a role of the thermodynamic critical temperature $T_{CN}$ for Cu copper metal with the first type order transition in magnetic field *B* at temperature $T < T^*$. The mean free path value $l \approx 1.3 \cdot 10^{-4}$ cm is in a good agreement with the magnitude of magnetic field *B*. There is an interesting situation in the Cu-Nb composite sample, where the full ultrasound attenuation value depends on the internal electron mean path in Cu copper matrix, and the proximity coherence length depends on the electron mean free path, which is dependent on the scattering of electron excitations on impurities in the Cu-Nb interfaces. The more rapid decrease of the ultrasonic longitudinal wave attenuation with the temperature decrease below 0.6 K (Fig. 6) can be connected with the partial correspondence of selected simplified theoretical model to the true multi-filaments structure in the researched Cu-Nb composite samples. At low temperature, the coherence length $\xi_N$ is large, and also there is an important long-distance order among the Nb filaments, which is not described by our prime model, and it can be necessary to proceed to a theoretical model of rectangular cells. In this case, the normal metal fraction decreases more rapidly with the increase of coherence length $\xi_N$ at low temperature.

We also researched the influence of low magnetic field *B* on the proximity effect properties in the above described Cu-Nb composite samples. In [7] it was shown that the proximity effect is very sensitive to the magnetic field of (2-3) Oe, and this magnetic field decreases the coherence length in ten times at low temperature. In our Cu-Nb composite samples, the magnetic field up to $H=100$ Oe doesn't impact the ultrasound attenuation, because the Cu-Nb composite samples have both the distinct structure of the Nb niobium filaments distribution in Cu copper matrix in comparison with the samples researched in [7], and there are the contacts between the Nb filaments in our Cu-Nb composite samples. The magnetic field doesn't penetrate into the Cu-Nb composite sample, because the system of Nb filaments has the percolation superconducting properties with the large value of superconducting critical current *J* [36]. We believe that the diamagnetic properties and screening effect are also present due to the existing su-perconducting *Nb* niobium filaments network in *Cu-Nb* composite samples.

## CONCLUSION.

We researched the influence by the proximity effect on the ultrasonic longitudinal wave attenuation in *Cu-Nb* composite samples, which consist of the *Cu* copper normal metal in close contact with the *Nb* superconductor at frequency of *30 MHz* at temperatures of *0.35–2 K*. We confirmed that the ultrasound longitudinal wave attenuation measurements is a most informative method for accurate characterisation of the *S-N-S* system physical properties in comparison with the electrical current measurement method. The electron mean free path *l*, which is connected with the impurity scattering by the electrons in volume and near the *Cu-Nb* interfaces in a *Cu-Nb* composite sample, was measured by the three independent methods at ultra low temperatures. It is assumed that the *Andreev reflections* were present in *S-N-S* structure, creating a certain influence on the measured dependence of ultrasonic longitudinal wave attenuation in the *Cu-Nb* composite samples at ultra low temperatures. The measured experimental results of ultrasonic longitudinal wave attenuation in the *Cu-Nb* composite samples at ultra low temperatures are in good agreement with the theoretical modeling data, obtained in the case of dirty local limit.

This article was published in Russian in Problems of Atomic Science and Technology (*VANT*) [37].

———————